\documentclass[preprint,
superscriDptaddress,
amsmath,amssymb,aps,showkeys,showpacs,
twoside,final,secnumarabic,
nofootinbib]{revtex4-2}

\usepackage[paperwidth=205mm,paperheight=290mm,top=17mm,bottom=25mm,
inner=17mm,outer=17mm,
twoside]{geometry}

\usepackage{cmap} 
\usepackage[T1,T2A]{fontenc}
\usepackage[utf8]{inputenc}
\usepackage[russian,english]{babel}
\usepackage{color}
\usepackage{graphicx}
\usepackage{dcolumn}
\usepackage{bm} 
\usepackage[unicode=true,colorlinks=true,linkcolor=magenta, urlcolor=blue, citecolor = blue,breaklinks]{hyperref}
\usepackage{multirow}
\usepackage{url}
\usepackage{breakurl}
\DeclareGraphicsExtensions{.eps}

\newcount\issue
\newcount\Vol
\newcount\numb
\usepackage{fancyhdr} 
\def\Vol{\textbf{80}}
\def\numb{x}
\begin{document}

\title{ 
Manifestations of Dark Matter \\ in processes with three and four Top-Quarks} 

\def\addressa{Skobeltsyn Institute of Nuclear Physics, Lomonosov Moscow State University, Leninskie gory, GSP-1}
\def\addressb{Sarov Branch of Lomonosov Moscow State University,\\
Sarov 607328, Russian Federation}

\author{\firstname{E.}~\surname{Abasov}}
\affiliation{\addressa}
\author{\firstname{E.}~\surname{Boos}}
\affiliation{\addressa}
\author{\firstname{V.}~\surname{Bunichev}}
\affiliation{\addressa}
\author{\firstname{P.}~\surname{Volkov}}
\affiliation{\addressa}
\author{\firstname{G.}~\surname{Vorotnikov}}
\affiliation{\addressa}
\author{\firstname{L.}~\surname{Dudko}}
\affiliation{\addressa}
\author{\firstname{A.}~\surname{Zaborenko}}
\affiliation{\addressa}
\author{\firstname{E.}~\surname{Iudin}}
\affiliation{\addressa}
\author{\firstname{A.}~\surname{Markina}}
\email[E-mail: ]{berezina@www-hep.sinp.msu.ru }
\affiliation{\addressa}
\author{\firstname{M.}~\surname{Perfilov}}
\affiliation{\addressa}
\author{\firstname{N.}~\surname{Savkova}}
\affiliation{\addressb}

\received{xx.xx.2025}
\revised{xx.xx.2025}
\accepted{xx.xx.2025}

\begin{abstract}
This study explores possible manifestations of Dark Matter (DM) in processes involving multiple top quarks at the LHC. We analyze both the associated production of scalar ($\phi$) and pseudoscalar ($a$) DM Mediators with up to four top quarks, as well as their resonant production in top-rich final states, within a simplified model framework. Cross sections were calculated using the CompHEP and MadGraph packages with benchmark parameters recommended by the LHC Dark Matter Working Group. A detailed comparison of differential cross sections for key kinematic observables in three- and four-top-quark production demonstrates pronounced deviations from Standard Model predictions. The results highlight the limited feasibility of associated production channels, while resonant DM Mediator production offers realistic discovery prospects. Overall, the analysis underscores the potential of multi-top-quark final states as sensitive probes of Dark Matter at current and future collider experiments.
\end{abstract}

\pacs{PACS: 12.60.-i, 14.65.Ha, 95.35.+d;}\par
\keywords{Keywords: Dark Matter, Top Quark, Simplified Models, LHC  \\[5pt]}

\maketitle
\thispagestyle{fancy}


\section{Introduction}\label{intro}

The gravitational evidence for non-luminous matter, collectively termed Dark Matter (DM), is one of the most robust conclusions of modern cosmology and astrophysics. This evidence is drawn from a multitude of independent observations, including the anomalous orbital velocities of stars in galaxies~\cite{Rubin:1970zza} and the dynamics of galaxy clusters, famously demonstrated through the Bullet Cluster merger, which shows a clear separation between luminous matter and the gravitational potential~\cite{Clowe:2006eq}. The consensus picture from these observations is that DM is cold, collisionless, and interacts at most weakly with the Standard Model (SM) sector.

A primary focus of particle physics has been to identify the microscopic nature of DM, with a leading hypothesis positing it as a new, weakly-interacting massive particle (WIMP). The WIMP paradigm is particularly compelling as it naturally yields the correct relic abundance through thermal freeze-out in the early universe~\cite{Jungman:1995df}. This framework motivates a diverse experimental program to produce or detect these particles. Direct detection experiments aim to observe the rare recoil of a DM particle scattering off a nuclear target~\cite{LUX:2016ggv}, while indirect detection searches look for SM products of DM annihilations in astrophysical regions of high density.

Collider experiments offer a complementary and powerful approach. The Large Hadron Collider (LHC) can probe DM production directly, potentially creating these particles in proton-proton collisions. The canonical signature is the presence of significant missing transverse momentum ($\vec{E}_T^{\text{miss}}$), imbalance indicating the escape of stable, neutral particles. This signal is typically accompanied by the initial-state radiation of a visible SM particle, such as a jet, photon, or weak boson, which serves to tag the event and trigger the detector~\cite{Bertone:2004pz}.

The interpretation of DM searches at the LHC is guided by two complementary theoretical frameworks. At energy scales significantly below the mass of a potential Mediator, an Effective Field Theory (EFT) description provides an appropriate parameterization. In this approach, DM interactions with quarks and gluons are described by non-renormalizable operators characterized by their mass scale and Lorentz structure~\cite{Goodman:2010ku}. However, the validity of this approximation becomes questionable when the collision energy approaches the Mediator mass scale—precisely the regime probed by the LHC.

For such scenarios, simplified models offer a more robust and general framework~\cite{Abdallah:2014hon}. These models extend the Standard Model through minimal additions, typically comprising a DM particle and an s- or t-channel DM Mediator with specified quantum numbers and interaction vertices. This renormalizable framework enables self-consistent calculations of production cross-sections and kinematic distributions, while facilitating direct comparison with cosmological constraints~\cite{Abercrombie:2015wmb}. The most thoroughly investigated implementations incorporate spin-0 (scalar/pseudoscalar) and spin-1 (vector/axial-vector) DM Mediators, each producing distinctive collider signatures that enable targeted experimental searches.

The framework of Minimal Flavor Violation (MFV)~\cite{Chivukula:1987py,Hall:1990ac,Buras:2000dm,DAmbrosio:2002vsn} suggests that third-generation quarks could play a dominant role in DM interactions~\cite{Lin:2013sca}. This naturally directs attention to collider processes where DM is produced in association with the top quark, the heaviest Standard Model particle. Consequently, extensive experimental searches have been conducted for final states containing DM alongside single top quarks ($t/\bar{t}$ + DM) or top quark pairs ($t\bar{t}$ + DM)~\cite{CMS:2019zzl,ATLAS:2022znu,ATLAS:2022ygn}. These channels provide exceptional sensitivity to new physics scenarios characterized by enhanced couplings to the heavy-quark sector. To date, no significant excess over Standard Model predictions has been observed at the LHC, underscoring the critical importance of continuously improving analysis techniques and search sensitivity.

In simplified DM models Dark Matter particles $\chi$ are Dirac fermions interacting with the SM and DM sectors via either a massive electrically neutral scalar DM Mediator $\phi$ or a pseudoscalar DM Mediator $a$ ~\cite{Buckley:2014fba}. The interaction Lagrangians of scalar and pseudoscalar particles with SM and DM fermions are as follows:

\begin{equation}
\begin{split}
L_{\phi} = g_\chi \phi \bar{\chi} \chi + \frac{g_f \phi}{\sqrt{2}} \sum_f \left(y_f \bar{f} f \right) \\
L_a = i g_\chi a \bar{\chi} \gamma^5 \chi + i \frac{g_f a}{\sqrt{2}} \sum_f \left(y_f \bar{f} \gamma^5 f \right)
\end{split}
\label{lagrangian}
\end{equation}

Here, summation is done over all SM fermions, denoted as $f$. The parameters $y_f = \sqrt{2} m_f / v$ are Yukawa coupling constants with the vacuum expectation value of the Higgs field being 246 GeV. $g_\chi$ is the coupling constant of DM fermions with the DM Mediator and $g_f$ is the coupling constant of SM fermions with the DM Mediator. it is usually assumed that the value of $g_\chi$ is the same for all fermion flavors. Under the assumption of minimal flavor violation, such a simplified model contains a minimal set of four free parameters: the mass of Dark Matter particles $m_\chi$, the mass of the DM Mediator $m_{\phi/a}$, $g_f$, and $g_\chi$. According to the recommendations of the LHC Dark Matter Working Group~\cite{Boveia:2016mrp} study the coupling parameter values are taken to be $g_f = g_\chi = 1$, and the DM particle mass $m_\chi = 1$ GeV in the paper.

The top quark's unique status as the Standard Model particle with the largest Yukawa coupling makes processes involving its interaction with (DM) particularly promising from both experimental and phenomenological perspectives. While single and pair top quark signatures have been extensively studied, triple top quark production \cite{Boos:2021yat} represents another compelling channel for DM searches at the LHC. This is complemented by four-top-quark production, a rare Standard Model process whose observation has recently been established by the LHC experiments~\cite{CMS:2023ftu, ATLAS:2023ajo}. The relevance of these multi-top final states for DM searches is underscored by considering DM Mediators with masses near current experimental limits of approximately 410 GeV~\cite{CMS:2025ncs,Stafford:2025boo,Chalbaud:2024jsr,Azevedo:2023xuc}. For such a Mediator, the branching fraction to a top quark pair can reach approximately 16\% for scalar DM Mediator and 28\% for pseudoscalar DM Mediator, suggesting that DM processes could substantially contribute to observed triple and four-top-quark event rates \cite{Abasov:2024mwk}.

In this paper, we investigate how scalar and pseudoscalar mediators could manifest through multi-top-quark final states, providing distinctive collider signatures beyond existing DM searches.
Thе paper is organized as follows. Section \ref{sec:2} examines the associated production of DM Mediators with single, double, triple, and four top quarks and assesses their detection prospects at the LHC. Section \ref{sec:3} investigates resonant DM Mediator production leading to two, three and four top quarks in the final state. Finally, Section \ref{sec:conclusion} summarizes the key findings and conclusions. 

All numerical calculations presented in this work were performed using the CompHEP package \cite{CompHEP:2004qpa,comphep:webpage} and the MadGraph event generator~\cite{Alwall:2014hca}. The values in the tables reflect only the Monte Carlo statistical uncertainty. The calculations of total cross-sections were performed using a fixed choice for the renormalization and factorization scales, which was set according to the characteristic energy scale of the specific process under study, while the default MadGraph dynamic scale was used for the distributions in the final section. The variation of the scales demonstrates about 10\% uncertainty. All cross-section calculations in this work were performed using the complete set of QCD and electroweak diagrams. This approach is particularly important for three- and four-top-quark production processes, where the interference between QCD and electroweak contributions is known to be substantial and destructive, significantly affecting the accuracy of theoretical predictions~\cite{Boos:2021yat, Abasov:2024mwk}. The calculations were performed for the LHC proton-proton collider at a center-of-mass energy of $\sqrt{s} = 14$ TeV.

\section{\label{sec:2} Associated Production of a Dark Matter Mediator with Top Quarks
}
In modern collider experiments, the primary direction in the search for Dark Matter is the associated production of DM with SM particles. In this section, we compare the potential contribution of DM with analogous processes within the SM. The calculations are performed with the same benchmark parameter values to compare the relative contribution of DM and estimate the overall sensitivity. 

\subsection{\label{sec:2_1} Processes of Associated Production of a Dark Matter Mediator with a Single Top Quark and a Top-Quark Pair}

\begin{figure}[!h!tb]
\includegraphics[width = .80\linewidth]{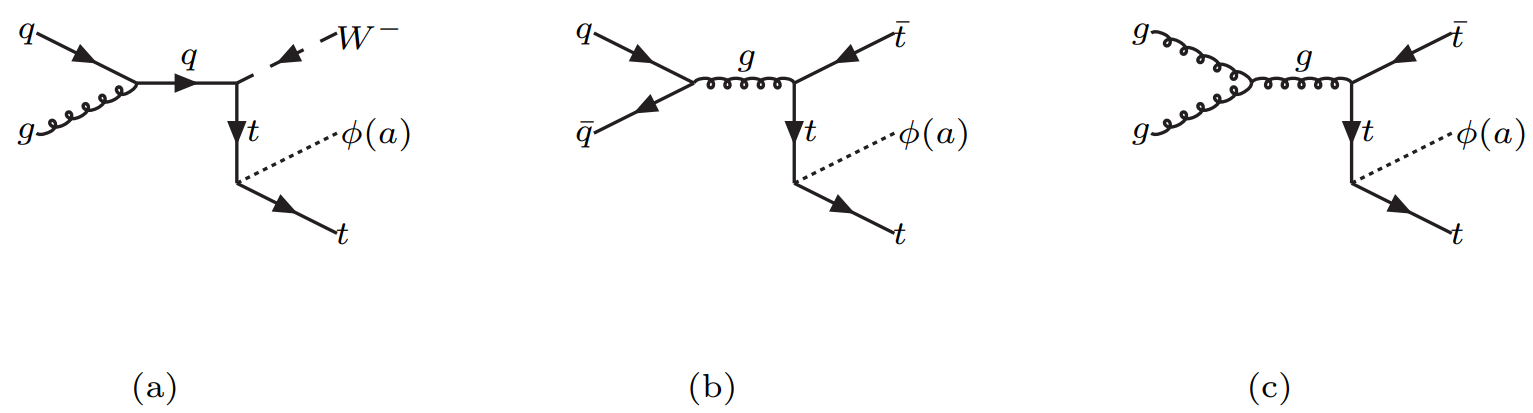}
\caption{\label{fig:assoc_diag} Representative diagrams for the associative production of a scalar and pseudoscalar DM Mediator with one (a) and two (b,c) top quarks.}
\end{figure}

As part of the study of possible processes of DM particle production in association with top quarks, the cross sections of the scalar and pseudoscalar DM Mediator associated production with single and double top quarks were calculated. They are listed in Tables ~\ref{tab:tWphi} and ~\ref{tab:ttphi}. Representative Feynman diagrams for these processes are shown in  Fig.~\ref{fig:assoc_diag}. Table~\ref{tab:tWbphi} presents the cross sections for the final state $t W^{-} \bar{b}\phi(a)$, which combines pair and single production of top quarks at  NLO in association with the DM Mediator.

\begin{table}[h]
        \centering
        \begin{tabular}{|c|c|}
            \hline
            \multicolumn{2}{|c|}{$t  W^- \phi (a)$ } \\ \hline
            Process  & Cross-section (fb) \\ \hline
            $bg \to t  W^- \phi$ & 2.38 $\pm( 9.80\times 10^{-4})$  \\ \hline
            $dg \to t  W^- \phi$ & $ 1.96\times 10^{-3} \pm (8.39\times 10^{-7})$  \\ \hline
            $sg \to t  W^- \phi$ & $ 1.12\times 10^{-2} \pm (4.62\times 10^{-6})$  \\ \hline
            $pp \to t  W^- \phi$ & $ 2.40 \pm (9.80\times 10^{-4})$  \\ \hline\hline
            $bg \to t  W^- a$ & 2.30 $\pm(1.82\times 10^{-3})$  \\ \hline
            $dg \to t  W^- a$ & $ 1.89\times 10^{-3} \pm (1.63\times 10^{-6})$  \\ \hline
            $sg \to t  W^- a$ & $ 1.08\times 10^{-2} \pm (8.71\times 10^{-6})$  \\ \hline
            $pp \to t  W^- a$ & $ 2.31 \pm (1.82\times 10^{-3})$  \\ \hline

        \end{tabular}
        \caption{Cross sections of the process $pp \to t  W^- \phi(a)$ and its subprocesses  within the simplified DM model with a scalar $\phi$ (pseudoscalar $a$) DM Mediator in CompHEP.}
        \label{tab:tWphi}
    \end{table}

\begin{table}[h]
    \centering
    \begin{tabular}{|c|c|}
        \hline
        \multicolumn{2}{|c|}{$t \bar{t} \phi (a)$} \\ \hline
        Process  & Cross-section (fb) \\ \hline
        $u \bar{u} \to t \bar{t} \phi$ & $1.48  \pm (1.31 \times 10^{-3}$) \\ \hline
        $d \bar{d} \to t \bar{t} \phi$ & $0.85  \pm (7.64 \times 10^{-4}$) \\ \hline
        $g g \to t \bar{t} \phi$ & $13.73  \pm (2.37 \times 10^{-2}$) \\ \hline 
        $p p \to t \bar{t} \phi$ & $16.61 \pm (2.38 \times 10^{-2}$) \\ \hline \hline
        $u \bar{u} \to t \bar{t} a$ & $0.58  \pm (3.37 \times 10^{-4}$) \\ \hline
        $d \bar{d} \to t \bar{t} a$ & $0.33  \pm (2.04 \times 10^{-4}$) \\ \hline
        $g g \to t \bar{t} a$ & $25.06  \pm (2.80 \times 10^{-2}$) \\ \hline
        $p p \to t \bar{t} a$ & $26.11 \pm (2.80 \times 10^{-2}$) \\ \hline 
    \end{tabular}
    \caption{Cross sections of the process $pp \to t \bar{t} \phi(a)$ and the main subprocesses within the simplified DM model with a scalar $\phi$ (pseudoscalar $a$) DM Mediator in CompHEP.}
    \label{tab:ttphi}
\end{table}

 \begin{table}[h]
    \centering
    \begin{tabular}{|c|c|}
        \hline
        \multicolumn{2}{|c|}{$t W^{-} \bar{b} \phi (a)$ } \\ \hline
        Process & Cross-section (fb) \\ \hline
        $u \bar{u} \to t W^{-} \bar{b} \phi$ & 1.52  $\pm( 8.00 \times 10^{-4})$ \\ \hline
        $gg \to t W^{-} \bar{b} \phi$ & 16.50 $\pm (4.71 \times 10^{-2})$ \\ \hline
        \hline
        $u \bar{u} \to t W^{-} \bar{b} a$ & 0.59  $\pm( 5.36 \times 10^{-3})$ \\ \hline 
        $gg \to t W^{-} \bar{b} a$ & 28.0 $\pm (6.61 \times 10^{-2})$ \\ \hline
    \end{tabular}
    \caption{Cross sections of some subprocesses of the process $pp \to t W^{-} \bar{b} \phi (a)$ within the simplified DM model with a scalar $\phi$ (pseudoscalar $a$) DM Mediator in CompHEP.}
    \label{tab:tWbphi}
\end{table}

A recent CMS experimental analysis of these processes reports an upper limit on the production cross-section of scalar and pseudoscalar DM Mediators with a mass of 400 GeV of 50 fb at the 95\% confidence level~\cite{CMS:2025ncs}. In comparison, the associated production of $tW\phi (a)$ (even with relatively high benchmark couplings) is approximately four orders of magnitude lower than the standard model $tW$ production. The potential contribution from Dark Matter to top-quark pair production is also negligible and statistically difficult to detect.

\subsection{\label{sec:2_2} Processes of Associated Production of a Dark Matter Mediator with three and four Top Quarks}

The Tables \ref{tab:3tphi} and \ref{tab:4tphi} shows the cross sections of the DM Mediator production processes together with three and four top quarks. Their values are 3-4 orders of magnitude smaller than the cross sections for the production of three and four top quarks in the Standard Model and are beyond the sensitivity of the LHC detectors at the current stage. Thus, the search for Dark Matter particles in these channels is unlikely in the foreseeable future. 

\begin{table}[h]
    \centering
     \begin{tabular}{|c|c|}
        \hline
        \multicolumn{2}{|c|}{$t \bar{t}t W^- \phi (a)$} \\ \hline
        Process  & Cross-section (fb)\\ \hline
        $pp \to t \bar{t} t W^- \phi$ &  2.10$\times10^{-3}$  $\pm$ (5.44$\times10^{-6}$) \\ \hline \hline
        $pp \to t \bar{t} t W^- a$ &   1.36$\times10^{-3}$  $\pm$ (3.54$\times10^{-6}$) \\ \hline
    \end{tabular}
        \caption{Cross sections of the process $pp \to t \bar{t}t W^- \phi (a)$ within the simplified DM model with a scalar $\phi$ (pseudoscalar $a$) DM Mediator in MadGraph.}
    \label{tab:3tphi}
\end{table}

\begin{table}[h]
    \centering
    \begin{tabular}{|c|c|}
        \hline
        \multicolumn{2}{|c|}{$t \bar{t}t \bar{t} \phi (a)$} \\ \hline
        Process  & Cross-section (fb)\\ \hline
        $pp \to t \bar{t} t \bar{t}\phi$ & 9.35$\times10^{-3}$   $\pm$ (2.80$\times10^{-5}$) \\ \hline \hline
        $pp \to t \bar{t} t \bar{t}a$ & 5.13$\times10^{-3}$   $\pm$ (1.47$\times10^{-5}$)  \\ \hline
       
    \end{tabular}
    \caption{Cross sections of the process $pp \to t \bar{t} t \bar{t} \phi (a)$ within the simplified DM model with a scalar $\phi$ (pseudoscalar $a$) DM Mediator in MadGraph.}
    \label{tab:4tphi}
\end{table}

\section{\label{sec:3}Resonant production of a Dark Matter Mediator in processes with top quarks
}
Another approach to Dark Matter searches in simplified models is through resonant Mediator production, where the DM Mediator decays to Standard Model particles. For the chosen benchmark parameters, the branching fraction of the scalar DM Mediator to a top-quark pair is approximately 16\%, and is even higher for the pseudoscalar case. These processes feature a fully reconstructible final state, offering significantly better sensitivity compared to associated DM production. In this section, we estimate the relative contribution of resonant  DM Mediator production, with subsequent decay to a top-quark pair, against the corresponding SM processes.

\subsection{\label{sec:3_1}Cross sections for resonant Dark Matter Mediator production into a final state of a top-quark pair} 
\begin{figure}[!h!tb]
\includegraphics[width = .15\linewidth]{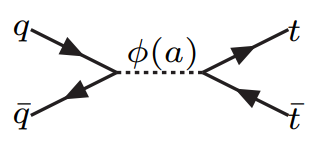}
\hspace{2cm} 
\includegraphics[width = .20\linewidth]{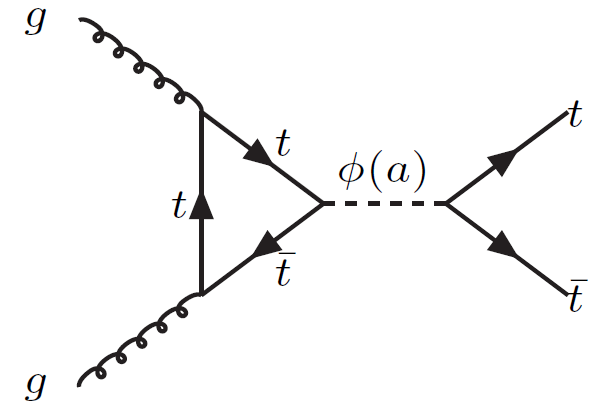}
\caption{\label{fig:res_diag} Representative diagrams for resonant production of a DM Mediator for two top-quark final states.}
\end{figure}

Processes of top pair production can be initiated by DM Mediator if the last one can decay to top quarks. At Fig.~\ref{fig:res_diag} two representative Feynman diagrams for the noted processes are provided. The first one is tree level diagram while the second one is the NLO loop diagram which is very important in this case. 

In Table~\ref{tab:2top_cx} the values of the cross sections for the processes of resonant DM production with two top quarks in the final state are provided. In comparison with the SM production of the top pair the DM Mediator contribution is tiny. If one considers tree level diagrams only the main contribution to the cross section comes from the subprocesses with two $b$-quarks in the initial state; however taking into account the NLO diagram with top loop increase the DM contribution to the cross section significantly.

\begin{table}[htb]
\centering
\begin{tabular}{|c|c|}
    \hline
    Process & Cross-section (pb)\\
    \hline
    $pp \to t\bar{t}$ (SM) & $598.6\pm (0.7)$\\  
    \hline \hline
    $pp \to t\bar{t}$ (SM + DM), scalar DM Mediator $\phi$& $595.9  \pm (0.7)$\\  
    \hline
    $q\bar{q} \to t\bar{t}$ (tree diagrams DM), scalar DM Mediator $\phi$& $0.005 \pm (0.005 \times10^{-3})$\\  
    \hline
    $gg \to t\bar{t}$ (loop diagram DM), scalar DM Mediator $\phi$& $0.6 \pm (0.6\times10^{-3})$\\ 
    \hline \hline
    $pp \to t\bar{t}$ (SM + DM), pseudoscalar DM Mediator $a$& $629.8  \pm (0.7)$\\  
    \hline
    $q\bar{q} \to t\bar{t}$ (tree diagrams DM), pseudoscalar DM Mediator $a$& $0.01 \pm (0.01\times10^{-3})$\\  
    \hline
    $gg \to t\bar{t}$ (loop diagram DM), pseudoscalar DM Mediator $a$ & $57.8  \pm (0.1)$\\ 
    \hline
    
\end{tabular}
\caption{Cross-sections for top quark pair production processes within the SM and within a simplified DM model with a scalar $\phi$ and pseudoscalar $a$ DM Mediator, calculated using MadGraph.}
\label{tab:2top_cx}
\end{table}

\subsection{\label{sec:3_2}Cross sections for resonant Dark Matter Mediator production into final states of three and four top quarks}

\begin{figure}[!h!tb]
\includegraphics[width = .80\linewidth]{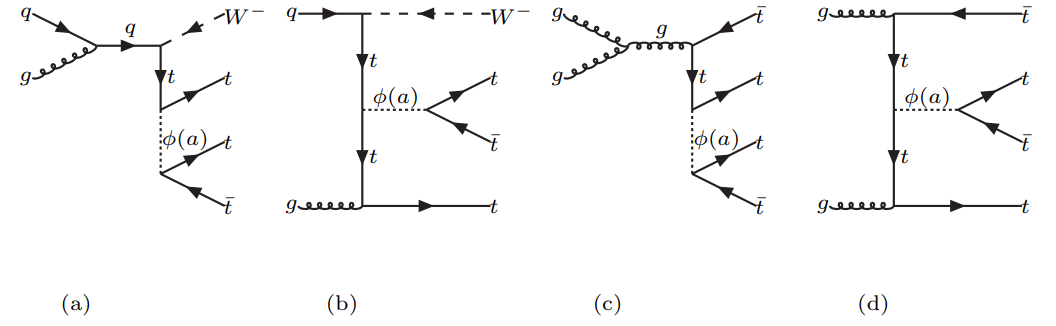}
\caption{\label{fig:res_diag_3_4t} Representative diagrams for resonant production of a DM Mediator  for three- (a,b) and four-top-quark (c,d) final states.}
\end{figure}

It has been shown~\cite{Abasov:2024mwk} that the scalar DM Mediator contribution to the total cross section of the processes with three or four top quarks can be significant. In this section the similar calculations are performed for the pseudoscalar DM Mediator and relative comparison in terms of the total cross sections are provided.
In Fig.~\ref{fig:res_diag_3_4t} the representative diagrams for the DM Mediator resonant production in association with one or two top quarks are presented. The DM Mediator then forced to decay into the pair of top quarks that leads to the final state that contains of three- or four-top quarks.
 
In Table~\ref{tab:3t_MG} one can see the cross section values for the processes with three top quarks and $W-$boson in the final state. The SM values and the contribution of the diagrams with resonant scalar and pseudoscalar DM Mediator are provided separately. One can conclude from the table values that the contribution of the DM Mediator resonant production to the total cross section is significant and can exceed the SM contribution, depending on the values of the coupling constants.

\begin{table}[!h!]
    \centering
    \resizebox{\textwidth}{!}{\begin{tabular}{|c|c|}
        \hline
        Process $pp \to t\bar{t}tW^-$ & MadGraph Cross-section (fb)\\
        \hline
        Full set of diagrams in the SM &$0.62 \pm (1.6\times10^{-3})$\\  
        \hline \hline
        Full set of diagrams in the DM model with a scalar DM Mediator $\phi$ & $1.54 \pm (5\times10^{-3})$\\  
        \hline
        Contribution of diagrams with a scalar DM Mediator $\phi$ &$0.90 \pm (2.4\times10^{-3})$\\  
        \hline \hline
        Full set of diagrams in the DM model with a pseudoscalar DM Mediator $a$ &$2.60 \pm (7.17\times10^{-3})$ \\  
        \hline
        Contribution of diagrams with a pseudoscalar DM Mediator $a$ & $1.98 \pm (5.77\times10^{-3})$ \\  
        \hline
      
    \end{tabular}}
    \caption{Cross-sections of the process $pp \to t\Bar{t}tW^-$ and individual contributions within the SM and the Simplified DM Model with a scalar(pseudoscalar)  DM Mediator. Full set of diagrams in the DM model comprises SM diagrams, diagrams involving the DM Mediator, and the interference between the SM and DM contributions.}
    \label{tab:3t_MG}
\end{table}

In Table~\ref{tab:4t_MG} the cross section values for the processes with four top quarks in the final state are listed. Again the SM values and the contribution of the diagrams with resonant scalar and pseudoscalar DM Mediator are provided separately. Contribution of the DM Mediator resonant production to the total cross section is significant but doesn't exceed the SM contribution with the same benchmark values of the couplings as for triple top-quark processes above. In addition one can observe the positive interference between set of DM diagrams and the SM ones, it can achieves~10\% of the total cross section, depending on the values of the coupling constants.

\begin{table}[!h!]
    \centering
    \resizebox{\textwidth}{!}{\begin{tabular}{|c|c|}
        \hline
        Process $gg \to t\Bar{t}t\Bar{t}$ & MadGraph Cross-section (fb)\\
        \hline
        Full set of diagrams in the SM &$7.79 \pm (2.3\times10^{-2})$\\  
        \hline \hline
        Full set of diagrams in the DM model with a scalar DM Mediator $\phi$ & $11.41 \pm (3.4\times10^{-2})$\\  
        \hline
        Contribution of diagrams with a scalar DM Mediator $\phi$ &$2.89 \pm (9.9\times10^{-3})$\\  
        \hline \hline
        Full set of diagrams in the DM model with a pseudoscalar DM Mediator $a$& $20.19 \pm (4.86\times10^{-2})$\\  
        \hline
        Contribution of diagrams with a pseudoscalar DM Mediator $a$ & $10.85 \pm (2.5\times10^{-2})$\\  
        \hline
      
    \end{tabular}}
    \caption{Cross-sections of the process $gg \to t\Bar{t}t\Bar{t}$ and individual contributions within the SM and the Simplified DM Model with a scalar(pseudoscalar) DM Mediator. Full set of diagrams in the DM model comprises SM diagrams, diagrams involving the DM Mediator, and the interference between the SM and DM contributions.}
    \label{tab:4t_MG}
\end{table}

The analysis examines the process $gg \to t\bar{t}tW^{-}\bar{b}$, a final state of special interest as it can originate from both three-top-quark production in association with a W-boson at NLO and four-top-quark production with subsequent decay of one top quark. Disentangling these contributions presents a methodological challenge analogous to the task of separating $tWb$ and $t\Bar{t}$ processes ~\cite{Boos:2023kpp}, further complicated by significant negative interference. The analysis of the complete set of diagrams for this process enables a comprehensive assessment of the DM contribution without requiring separate treatment of these channels.

Table~\ref{tab:3twb_MG} presents the cross-section values for the $gg \to t\bar{t}tW^{-}\bar{b}$ process calculated within the SM and the considered DM model. The results demonstrate that the contribution of DM Mediator diagrams to the $gg \to t\bar{t}tW^{-}\bar{b}$ cross-section is comparable to their contribution to the $gg \to t\bar{t}t\bar{t}$ process (Table~\ref{tab:4t_MG}), underscoring the dominant role of this channel.
\begin{table}[!h!]
    \centering
    \resizebox{\textwidth}{!}{\begin{tabular}{|c|c|}
        \hline
        Process $gg \to t\Bar{t}tW^-\Bar{b}$ & MadGraph Cross-section (fb)\\
        \hline
        Full set of diagrams in the SM & $15.55 \pm (4.7\times10^{-2})$\\  
        \hline \hline
        Full set of diagrams in the DM model with a scalar DM Mediator $\phi$ & $23.02 \pm (6.2\times10^{-2})$\\  
        \hline
        Contribution of diagrams with a scalar DM Mediator $\phi$ & $6.03 \pm (1.2\times10^{-2})$\\  
        \hline \hline
        Full set of diagrams in the DM model with a pseudoscalar DM Mediator $a$& $40.63 \pm (0.10)$\\  
        \hline
        Contribution of diagrams with a pseudoscalar DM Mediator $a$ & $22.27 \pm (5.17\times10^{-2})$\\  
        \hline
      
    \end{tabular}}
    \caption{Cross-sections of the process $gg \to t\Bar{t}tW^-\Bar{b}$ and individual contributions within the SM and the Simplified DM Model with a scalar(pseudoscalar) DM Mediator. Full set of diagrams in the DM model comprises SM diagrams, diagrams involving the DM Mediator, and the interference between the SM and DM contributions.}
    \label{tab:3twb_MG}
\end{table}

As the data in Tables~\ref{tab:3t_MG} - ~\ref{tab:3twb_MG} show, the contribution of diagrams with a DM Mediator can significantly enhance the total cross section compared to the SM prediction. It is important to note, however, that the presented cross-section values do not account for systematic uncertainties, such as the dependence on the choice of factorization and renormalization scales or the parametrization of parton distribution functions. Taking these uncertainties into account is essential for a quantitative interpretation of the results and may affect the precise assessment of the significance of the DM contribution.

\subsection{\label{sec:3_3}Comparison of Differential Cross Section Distributions between the Standard Model and a Dark Matter Model}

The search for DM particles in top-quark processes faces obvious difficulties at all levels, from theoretical predictions to experimental limitations. One of the main challenges is the need to separate the missing transverse momentum arising from undetectable Dark Matter particles from the missing transverse momentum of neutrinos~\cite{Abasov:2024nec}. For resonance Dark Matter production processes with three or four top quarks, such a procedure is not required, since Dark Matter particles do not appear in the final state in such processes. However, it is necessary to distinguish processes with three or four top quarks in the final state (signal process), which occur with a DM Mediator, from processes in the SM (background process). 
As it was shown in Sec.~\ref{sec:3_2}, the presence of scalar or pseudoscalar DM Mediators significantly changes the total cross section of three and four top quark production processes. The key observable for separating signal and background processes is the invariant mass of top-antitop pairs. In the signal DM process, where a Mediator with a mass of 400 GeV decays into a top-antitop pair, one of the possible pair combinations should exhibit a resonant peak around 400 GeV. For each event, all possible top-antitop pair combinations are considered, since it is not known a priori which pair originated from the Mediator decay.

\begin{figure*}[!h!tb]
\includegraphics[width = .45\linewidth]{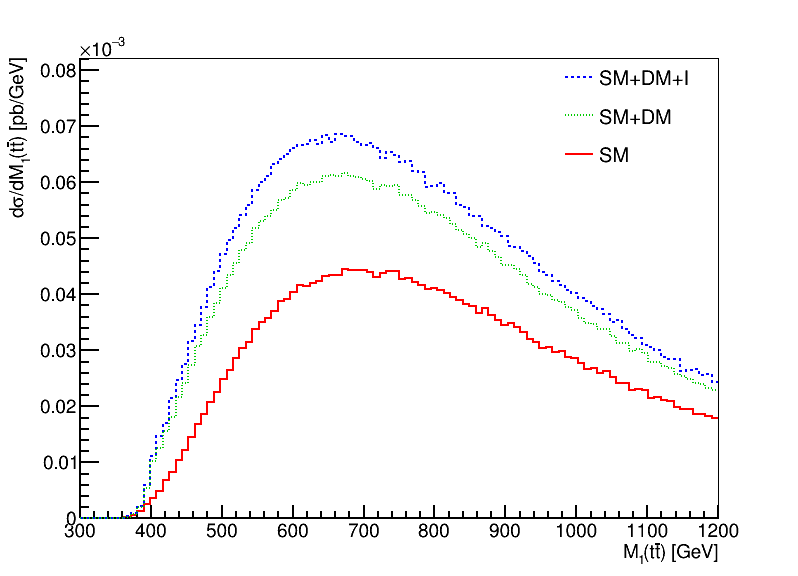}
\includegraphics[width = .45\linewidth]{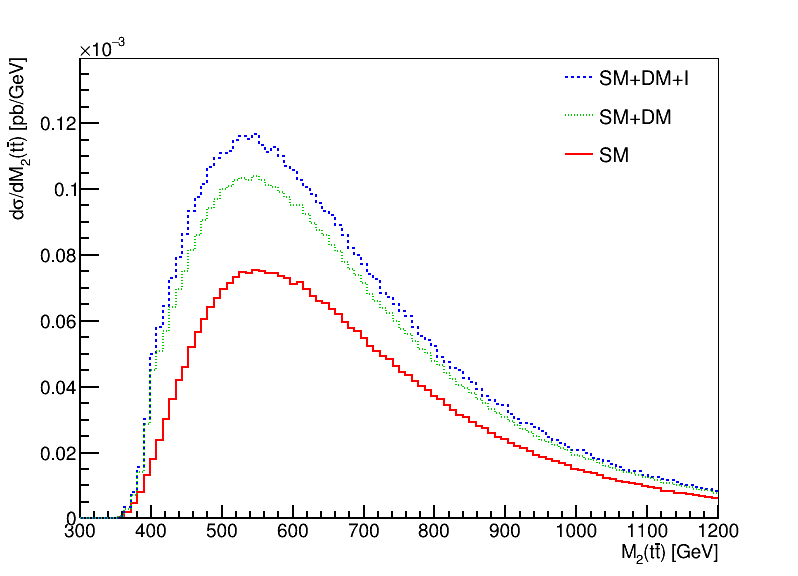}
\includegraphics[width = .45\linewidth]{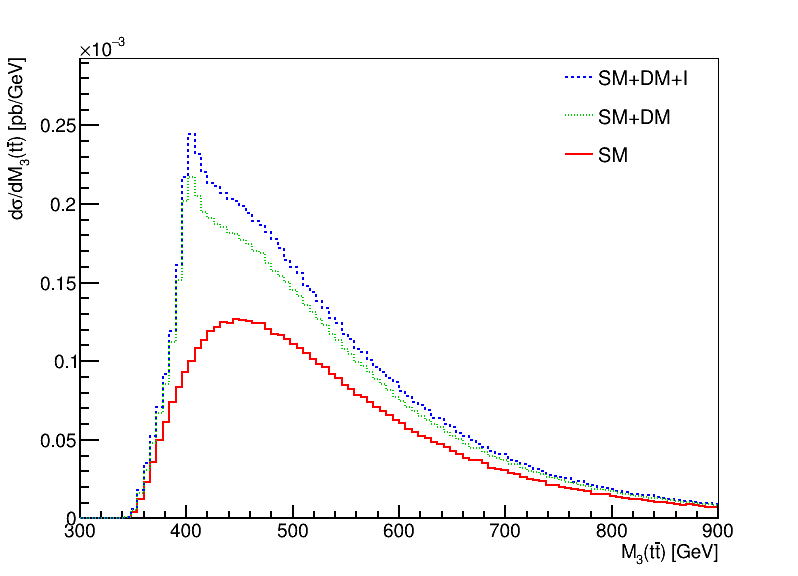}
\includegraphics[width = .45\linewidth]{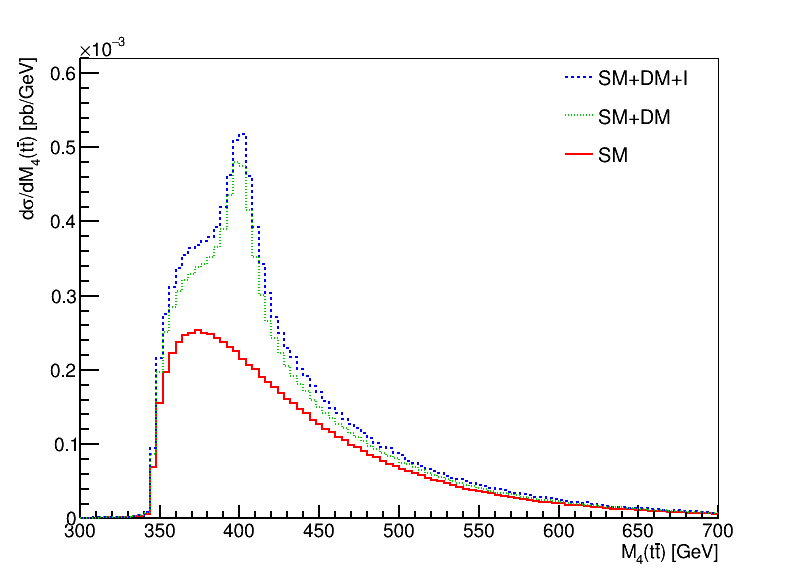}
\caption{\label{fig:dXsec_ttbar_mass} Distributions of invariant masses for all possible top-antitop pairs in the process $gg \to t\bar{t}tW^{-}\bar{b}$, ordered in descending order: $M_1(t\bar{t})$  (largest mass), $M_2(t\bar{t})$ , $M_3(t\bar{t})$ , and $M_4(t\bar{t})$ (smallest mass). The Standard Model process (red curve), the combined contribution of the Standard Model and the Dark Matter model with a scalar DM Mediator (green curve), and the full process including interference effects between the Standard Model and Dark Matter (blue curve).}
\end{figure*}

Fig.~\ref{fig:dXsec_ttbar_mass} shows the distributions of invariant masses for all possible top-antitop pairs in the process $gg \to t\bar{t}tW^{-}\bar{b}$, ordered in descending order: $M_1(t\bar{t})$  ($\text{Max } M(t\bar{t})$ - largest mass), $M_2(t\bar{t})$ , $M_3(t\bar{t})$ , and $M_4(t\bar{t})$ ($\text{Min } M(t\bar{t})$ - smallest mass). The presented distributions correspond to three different scenarios: the Standard Model process (red curve), the combined contribution of the Standard Model and the Dark Matter model with a scalar DM Mediator (green curve), and the full process including interference effects between the Standard Model and Dark Matter (blue curve). As can be seen from the distributions, the signal processes show a resonant peak at 400 GeV in the $M_3(t\bar{t})$  and $M_4(t\bar{t})$ distributions, which aligns with the expected behavior for the decay of a scalar DM Mediator. This peak is entirely absent in the pure SM background process, where the distributions are smooth and continuous without resonant features. It is noteworthy that no resonant peak is observed in the $M_1(t\bar{t})$  and $M_2(t\bar{t})$ distributions, indicating that the pair from the Mediator decay is rarely the heaviest in the event. Additionally, the negligible contribution of interference effects is evident from the close agreement between the distributions for the combined SM and DM contribution and the full process including interference.

\begin{figure}[!h!tb]
\includegraphics[width = .32\linewidth]{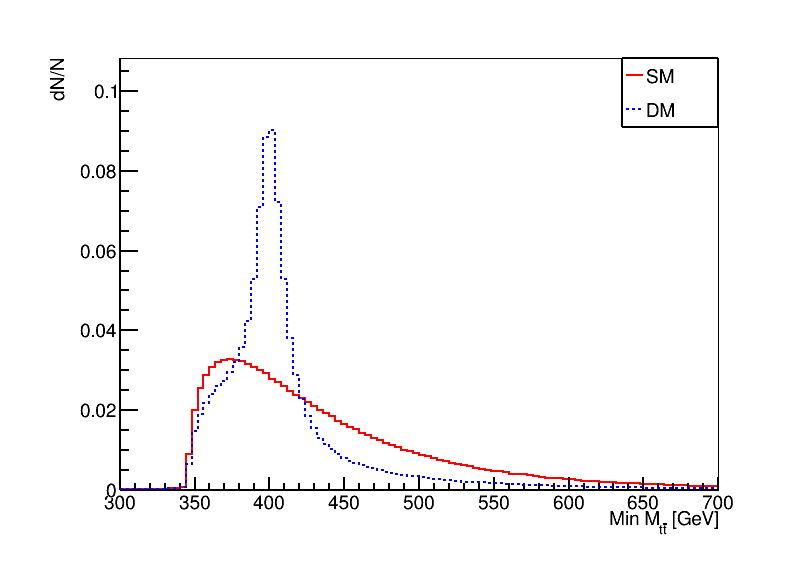}
\includegraphics[width = .32\linewidth]{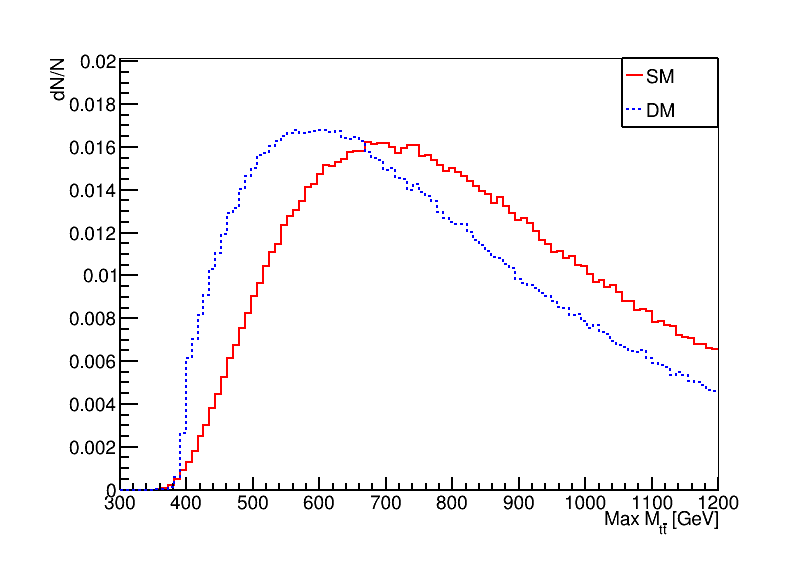}
\includegraphics[width = .32\linewidth]{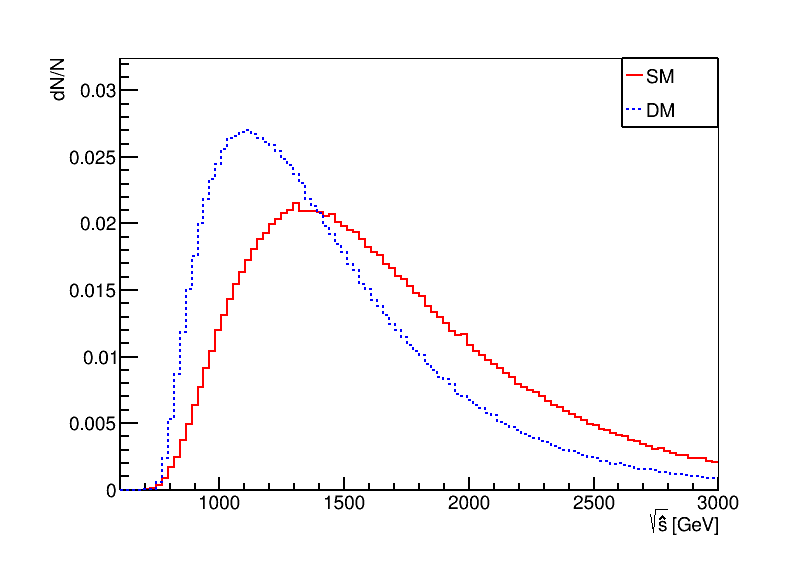}
\caption{\label{fig:form_comparison1} Normalized differential cross sections for $\text{Min } M(t\bar{t})$, $\text{Max } M(t\bar{t})$, $\hat{s}$. The Standard Model process (red curve), the Dark Matter model with a scalar DM Mediator (blue curve).}
\end{figure}

\begin{figure}[!h!tb]
\includegraphics[width = .32\linewidth]{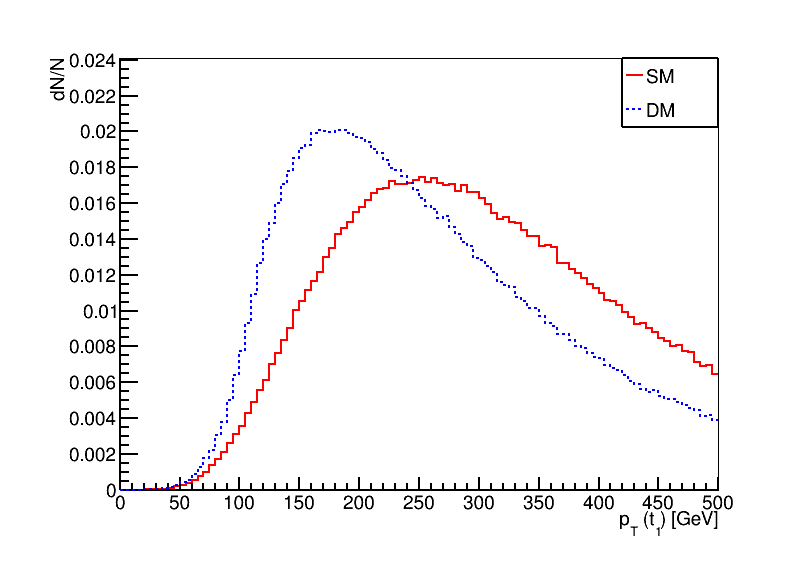}
\includegraphics[width = .32\linewidth]{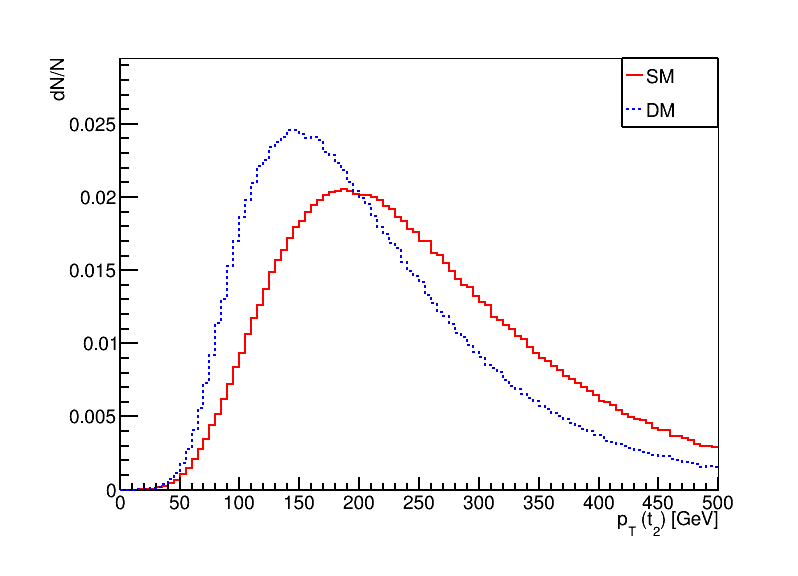}
\includegraphics[width = .32\linewidth]{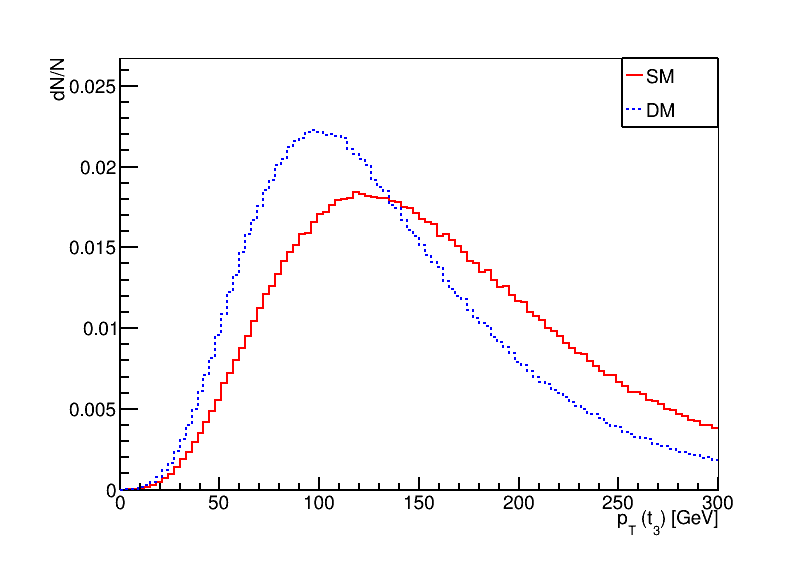}
\includegraphics[width = .32\linewidth]{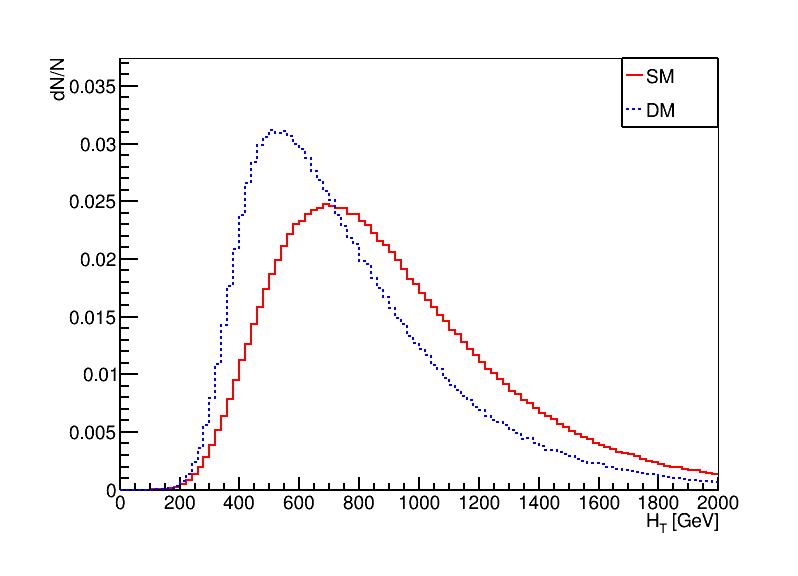}
\includegraphics[width = .32\linewidth]{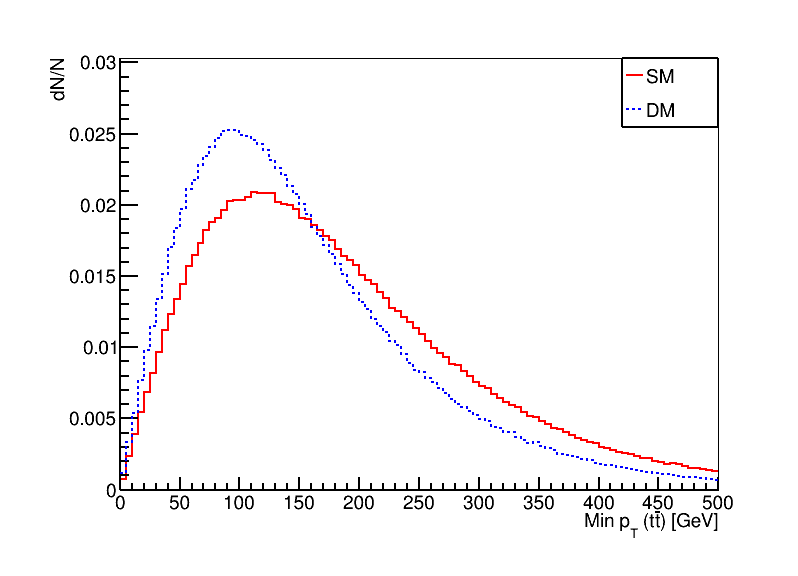}
\includegraphics[width = .32\linewidth]{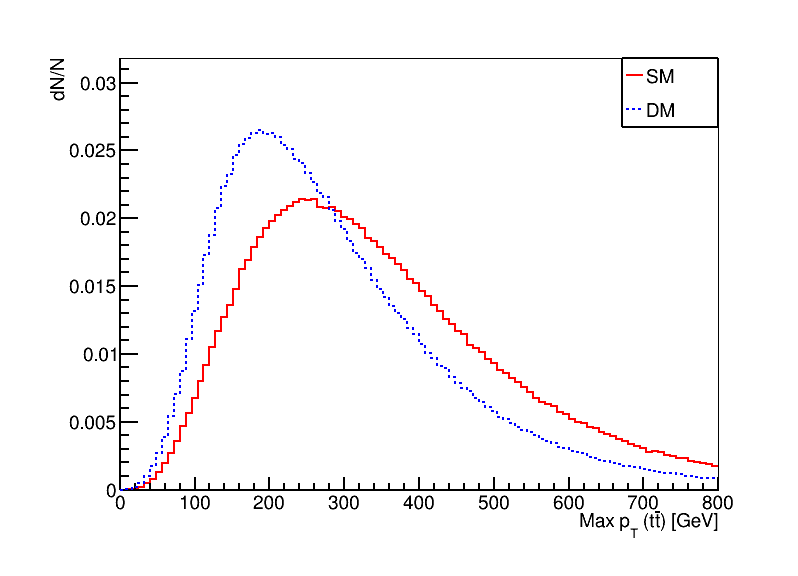}
\caption{\label{fig:form_comparison2} Normalized differential cross sections for $p_T(t_1)$, $p_T(t_2)$, $p_T(t_3)$, $H_T$, $\text{Min } p_T(t\bar{t})$, $\text{Max } p_T(t\bar{t})$. The Standard Model process (red curve), the Dark Matter model with a scalar DM Mediator (blue curve).}
\end{figure}

In addition, we present several other observables that exhibit different kinematic behavior for Standard Model and Dark Matter processes. Fig.~\ref{fig:form_comparison1}-\ref{fig:form_comparison2} shows a comparison of the shapes of normalized differential cross sections for these variables. The distributions include the $\text{Min } M(t\bar{t})$ (minimal invariant mass of the different resonant $t\bar{t}$ pairs). This variable serves as a direct indicator of the Mediator mass and is expected to be a powerful discriminator. To characterize the global event properties, we present the invariant mass of the final particles $\hat{s}$ and the scalar sum of the transverse momenta of all final-state particles $H_T$. We also show the transverse momenta of the three hardest top quarks $p_T(t_1)$, $p_T(t_2)$ and $p_T(t_3)$, as well as the minimum and maximum transverse momentum among the $t\bar{t}$ pairs: $\text{Min } p_T(t\bar{t})$ and $\text{Max } p_T(t\bar{t})$. From the Figures one can see that the presence of DM Mediator in the processes significantly changes the shapes and the behavior of the curves in comparison to the SM case. Therefore, the sensitivity can be significantly improved with multivariate analysis methods.

\section{\label{sec:conclusion}Conclusion}
In this work, we performed a phenomenological study of Dark Matter manifestations in processes with three and four top quarks at the LHC within simplified models with scalar and pseudoscalar DM Mediators. It was shown that the associated production of a DM Mediator with multiple top quarks has extremely small cross sections and remains beyond the current experimental sensitivity. At the same time, resonant production channels, where the Mediator decays into a top–antitop pair, demonstrate significantly higher prospects for observation.

The calculated cross sections for resonant DM Mediator production in final states with three and four top quarks reveal that the Dark Matter contribution can be comparable to or even exceed the Standard Model background, depending on Mediator type and coupling values. Moreover, the presence of positive interference effects with Standard Model amplitudes further enhances the total cross section in the four-top channel.

Differential cross-section distributions for key kinematic variables indicate distinctive signatures of Dark Matter processes, such as resonant structures in top–antitop invariant masses. These observables, together with an extended set of event-shape variables, can form the basis for powerful multivariate analyses aimed at disentangling Dark Matter signals from Standard Model backgrounds.

The obtained results emphasize the importance of multi-top-quark final states as a promising direction in collider Dark Matter searches. They can serve as a foundation for the optimization of future experimental strategies, helping to design more sensitive analyses in ongoing and upcoming runs of the LHC and future collider projects.

\begin{acknowledgments}
This study was conducted within the scientific program of the Russian National Center for Physics and Mathematics, section 5 «Particle Physics and Cosmology». 
\end{acknowledgments}
\clearpage


\end{document}